\documentclass[pra,twocolumn,superscriptaddress,nofootinbib]{revtex4}

\usepackage{color}
\usepackage{epsfig}

\newcommand{\be}{\begin{equation}}
\newcommand{\ee}{\end{equation}}
\newcommand{\ba}{\begin{array}}
\newcommand{\ea}{\end{array}}

\newcommand{\bea}{\begin{eqnarray}}
\newcommand{\eea}{\end{eqnarray}}
\newcommand{\beas}{\begin{eqnarray*}}
\newcommand{\eeas}{\end{eqnarray*}}
\newcommand{\n}{\nonumber}
\newcommand{\ka}{\kappa}
\newcommand{\eps}{\epsilon}
\newcommand{\lam}{\lambda}

\newcommand{\f}{\frac}

\begin{document}

\title{Generalized Dirac Oscillators with position-dependent mass}

\author{C.-L. Ho
}
\affiliation{Department of Physics, Tamkang University, Tamsui
251, Taiwan, R.O.C.}

\author{P. Roy}
\affiliation{Atomic~Molecular~and~Optical~Physics~Research~Group, Advanced Institute of Materials Science, Ton Duc Thang University, Ho~Chi~Minh~City, Vietnam}
\affiliation{Faculty of Applied Sciences, Ton Duc Thang University, Ho Chi Minh City, Vietnam}



\begin{abstract}
We study the $(1+1)$ dimensional generalized Dirac oscillator  with a position-dependent mass. In particular, bound states with zero energy as well as non zero energy have been obtained for suitable choices of the mass function/oscillator interaction. It has also been shown that  in the presence of an electric field, bound states exist if the magnitude of the electric field does not exceed a critical value. 
\end{abstract}



\maketitle

\section{ Introduction.} 

Dirac oscillator (DO) \cite{moh} is one of a very few relativistic systems which admits exact solutions. After it was introduced, the DO and its various generalizations have been widely studied \cite{do1,do2,do3,do4,do5,do6}. One of the areas where the DO has found most applications is quantum optics. It has been shown that the Jaynes-Cummings and some of their generalized versions can be related to the DO system \cite{jc1,jc2,jc3,jc4}. Recently, the DO in $(1+1)$ dimensions has also been realized experimentally \cite{expt},  and a proposal for experimentally observing a  two dimensional DO system in a single trapped ion has also been put forward \cite{jc2}. A recent interesting paper \cite{longhi} considers a classical wave analogue of the DO based on light propagation in engineered fiber Bragg gratings. In this scenario the mass in DO becomes position dependent and is related to the refractive index. In this context we would like to mention that the lower dimensional Dirac equation with a position dependent mass has also found applications in electronic optics in graphene \cite{kats}. In this work our objective is obtain solutions of the DO with a position dependent mass with/without an electric field. It will be seen that the effect of the electric field is to destroy the bound states and there exists a critical field beyond which bound states do not exist. Another aspect which we shall explore is the possibility of finding zero energy states. It will be shown that for a number of physically acceptable mass functions, exact zero energy states can be found.


\section{The Model and SUSY form}

The DO in the presence of a position dependent mass and a potential is given by (in units of $c=\hbar=1$)
\be
\left[\sigma_x (p_x -i\sigma_z f(x)) + \sigma_z m(x) + V(x) -  E\right]\,\psi (x)= 0,
\label{dirac1}
\ee
where $\psi=\left(\ba{c}\psi_1\\ \psi_2\ea\right)$.

As it is difficult to find exact solutions of Eq.\,(\ref{dirac1}) for general functions of $f(x), m(x)$ and $V(x)$, so in this paper  we shall restrict ourselves to the following situations. 

Firstly, we consider the case in which these three functions are all proportional to a function $W(x)$, i.e., 
\be
f(x)=\ka_f W(x),~~ m(x)=\ka_m W(x),~~ V(x)=\ka_v W(x),
\label{ansatz}
\ee
where $\ka_f, \ka_m$ and $\ka_v$ are real constants \cite{do2,tkachuk}.
In this case, the Dirac equation can be cast into a supersymmetric (SUSY) quantum mechanical form, and hence exact solutions can be obtained by matching $W(x)$ with a suitable superpotential of the well-known SUSY models as classified in  \cite{khare}.  General idea of this connection is discussed in this section, and some representative examples presented in Sect. III and IV.A.  

Secondly,  we shall consider Eq.(\ref{dirac1})  without  the constraint (\ref{ansatz}).  In this general case exact solutions for all energies are difficult to obtain. Noting the fact that zero energy states are of interest in condensed matter systems such as graphenes \cite{z1,z2,z3,z4,z5,z6,z7,z8,z9,z10,z11}, we shall therefore investigate possible zero energy solutions of Eq.\,(\ref{dirac1})  in Sect. IV.B and V.
 
Eq.(\ref{dirac1})  can be written as

\be
[\sigma_x p_x -\sigma_y f(x)+ \sigma_z m(x)  + V(x)-  E]\,\psi(x)= 0.
\label{dirac1a}
\ee
It is seen that the component spinors are entangled. To disentangle the components, we make the ansatz
\cite{do2}
\be
\psi(x)=[\sigma_x p -\sigma_y f(x)+ \sigma_z m(x) -V(x)+ E] {\tilde\psi}.
\ee
This leads to
\bea
\left[p_x^2 + f^2(x) +m^2  - (E-V(x))^2\right.\n\\
\left. -\sigma_z f^\prime(x)
-\sigma_y m^\prime(x)+ i\sigma_x V^\prime(x) \right]\,\tilde\psi(x)=0,
\label{dirac2}
\eea
where prime denotes derivative with respect to $x$. 

With Eq. (\ref{ansatz}),  we have
\bea
\left[p_x^2 + \left(\ka_f^2  +\ka_m^2  -\ka_v^2\right)\,W^2 + 2\ka_vEW \right.\n\\
\left. -\left(\sigma_z \ka_f + \sigma_y \ka_m - i\sigma_x \ka_v\right)\,W^\prime  \right]\,\tilde\psi=E^2 \tilde\psi.
\eea
The eigenvalue problem of the matrix $(\sigma_z \ka_f + \sigma_y \ka_m - i\sigma_x \ka_v)$ is
\be
\left(\sigma_z \ka_f + \sigma_y \ka_m - i\sigma_x \ka_v\right)\chi=\lam\,\chi.
\ee
The eigenvalues can be obtained from the secular equation
\bea
\left|
\begin{array}{cc}
\ka_f -\lam & -i(\ka_m+\ka_v)\\
i(\ka_m-\ka_v) & -\ka_f - \lam
\end{array}
\right|=0,
\label{secular}
\eea
and are given by
 $\lam=\sigma\sqrt{\ka_f^2  +\ka_m^2  -\ka_v^2}, \sigma=\pm 1$.
If we set 
\be
\tilde\psi(x)=\chi\phi(x),
\label{chi}
\ee
then $\phi(x)$ satisfies
\bea\label{susy}
&&\left[p_x^2 + \left(\ka_f^2  +\ka_m^2  -\ka_v^2\right)\,W^2 
 -\sigma\sqrt{\ka_f^2  +\ka_m^2  -\ka_v^2}\,W^\prime \right.\n\\&&~~~~~~\left.+ 2\ka_vEW \right]\,\phi=E^2 \phi.
\eea

Note that Eq.(\ref{susy}) is in the SUSY quantum mechanical form, i.e.,
\be
\left(p_x^2 + \tilde W^2 -\sigma \tilde W^\prime \right)\,\phi=\eps\phi,
\label{susy2}
\ee
with the superpotential
\be
\tilde W(x)=\sqrt{\ka_f^2  +\ka_m^2  -\ka_v^2}\,\left( W(x)+ \frac{\ka_v\, E }{\ka_f^2  +\ka_m^2  -\ka_v^2}\right),
\ee
and
\be
\eps=\frac{\ka_f^2  +\ka_m^2 }{\ka_f^2  +\ka_m^2  -\ka_v^2}\,E^2.\label{eps}
\ee
So one can find solution corresponding to the energy $E$ by matching Eq.(\ref{susy})  with various SUSY models, for example as 
listed in \cite{khare}.

From Eq.(\ref{susy}) we see that for  $|\kappa_v|> \pm \sqrt{k_m^2+k_f^2}$ bound states do not exist as the potential becomes complex. The quantity $\sqrt{k_m^2+k_f^2}$ is the critical strength of the electric field.

\section{Examples}
We shall now examine a number of examples. 
The situation where the mass term  $m(x)$ of Eq.(\ref{dirac1}) is constant has been considered before \cite{tkachuk}, so we shall not discuss this case here. Below we consider two situations: $V(x)=0$, and $V(x)\neq 0$, while $f(x), m(x)\neq 0$.

\subsection{$V(x)=0$}

In this case $\ka_v=0$, $\eps=E^2$, and 
\be
\tilde W(x)=\ka\, W(x),~~\ka=\sqrt{\ka_f^2  +\ka_m^2}.
\label{ka}
\ee

Exactly solvable models can be obtained simply by matching  Eq.(\ref{dirac1}) with appropriate SUSY models \cite{khare}.  Some examples are as follows.


\subsubsection*{$\bullet$ Scarf II model}

Take

\be\label{scarfII}
\tilde W=A\tanh x + B \,{\rm sech}\, x.
\ee
Then
\be
E^2_\sigma=A^2-(A-n_\sigma)^2,
\ee
with
$n_+=0,1,2,\ldots, n_{\rm max}<A$ for $\sigma=1$, and  $n_-=1,2,\ldots, n_{\rm max}<A$ for $\sigma=-1$

\subsubsection*{$\bullet$ Rosen-Morse II model}

Take

\be
\tilde W=A\tanh x + \frac{B}{A}, ~~ B< A^2.
\label{RM}
\ee

Then
\be
E^2_\sigma=A^2 + \frac{B^2}{A^2}-(A-n_\sigma)^2 - \frac{B^2}{(A-n_\sigma)^2},
\label{E-RM}
\ee
with
\bea
n_+&=&0,1,2,\ldots, n_{\rm max}<A,~~  \sigma=1,\n\\
n_-&=&1,2,\ldots, n_{\rm max}<A, ~~\sigma=-1.
\label{n-RM}
\eea

It is not difficult to see that for $\kappa_v=0$, there are many other profiles of $f(x)$ and $m(x)$ for which the system admits exact solutions \cite{khare}. In the next section we shall consider the case in which $V(x)\neq 0$.


\subsection{$V(x)\neq 0$}

For this case we mention two representative examples below.

\subsubsection*{$\bullet$ harmonic oscillator}

The case with $f(x)=\ka_f x, V(x)=\ka_v x$ and $m(x)=m+\ka_m x$ ($m=$constant) was studied in \cite{do2}.

\subsubsection*{$\bullet$ Rosen-Morse II model}

If we take
\be
W(x)=\alpha_0 \tanh x,
\ee
then
\be
\tilde W(x)=\ka\left(\alpha_0 \tanh x +\frac{\ka_v E}{\ka^2}\right).
\label{RM2}
\ee
Here $\ka$ is given in Eq.(\ref{ka}).

Comparing Eq.(\ref{RM2}) with Eq.(\ref{RM}),  we have
\be
A=\alpha_0\ka, ~~B=\alpha_0\ka_v E.
\ee
Then from Eqs.(\ref{eps}) and (\ref{E-RM}), we find
\be
E_\sigma^2=\frac{\alpha_0^2\ka^2-(\alpha_0\ka-n_\sigma)^2}{1+\frac{\alpha_0^2\ka^2}{(\alpha_0\ka-n_\sigma)^2}
},
\ee
where $n_\pm$ are given by Eq.(\ref{n-RM}).


\section{Zero energy states}

Quite often it becomes impossible to find exact analytical solutions of all energy values. It may however be possible to find the exact analytical zero energy state.  In recent years zero energy states are of considerable interest in condensed matter physics, particularly in systems related to graphenes \cite{z1,z2,z3,z4,z5,z6,z7,z8,z9,z10,z11}.   Here our objective is to show that for a number of physically acceptable mass function the zero energy states can be found exactly.

\subsection{Case I}

Eq.(\ref{dirac1a}) with $E=0$ is
\be
\partial_x\psi(x)=-\left(\sigma_z f(x) + \sigma_y m(x) + i\sigma_x V(x)\right)\psi(x).
\ee
With the ansatz (\ref{ansatz}) it becomes
\be
\partial_x\psi(x)=-\left(\sigma_z \ka_f + \sigma_y \ka_m + i\sigma_x \ka_v\right)\,W(x)\,\psi(x).
\ee
Again we assume the solution to have the form
\be
\psi(x)=\chi\,\phi(x),
\ee
where the spinor $\chi$ satisfies
\be
\left(\sigma_z \ka_f + \sigma_y \ka_m + i\sigma_x \ka_v\right)\chi=\lam\,\chi.
\ee
The eigenvalues are 
 $\lam_\sigma=\sigma\sqrt{\ka_f^2  +\ka_m^2  -\ka_v^2}, \sigma=\pm 1$, and the corresponding eigenfunctions are
 \bea
 \chi_\sigma\sim 
 \left(
 \begin{array}{c}
   1 \\
  i \frac{\lam_\sigma- \ka_f }{\ka_m-\ka_v}
  \end{array}
  \right).
 \eea
$\phi_\sigma(x)$ is solved from the equation
 \be
 \partial_x\phi_\sigma(x)=-\lam_\sigma W(x)\phi_\sigma(x),
 \ee
 whose solution is
 \be\label{zero}
\phi_\sigma(x)\sim e^{-\lam_\sigma\int^x dx W(x)}.
\ee
For the $E=0$ solution, one should choose $\lam_\sigma$ and $W(x)$ such that $\phi_\sigma(x)$ is a square integrable function.

As an example, let us consider a generalization of the Rosen-Morse II potential and take  
\be
W(x)
=\tanh^{2n+1} x+\mu,~~n=0,1,2,\cdots,
\ee
where $\mu$ is a constant.
In this case the zero energy solution is given by
\bea
\label{zero1}
&&\phi_+(x)\sim \exp\left\{-\lam_+ \left[\mu x \right.\right.\\
&&\left.\left. +\f{1}{2n+1} {_2 F_1}(1,n+1,2+n,\tanh^2x)~\tanh^{2n+2}x\right]\right\}.\n
\eea

We would like to point out that it is possible to find many other examples, e.g, a generalization of (\ref{scarfII}) for which zero energy states can be found analytically.


\subsection{Case II}
As the second case, we consider Eq.\,(\ref{dirac1a}) with $V(x)=0$ , i.e.,
\be
[\sigma_x p_x -\sigma_y f(x)+ \sigma_z m(x)  -  E]\,\psi(x)= 0.
\label{dirac1b}
\ee
We now assume the profile of $f(x)$ and $m(x)$ to be
\bea
f(x) \equiv \left\{
               \begin{array}{ll}
                f_+, & x\geq 0;\\
               - f_- , & {\rm otherwise}.
                        \end{array}
               \right.
\label{f}
\eea
and
\bea
m(x) \equiv \left\{
               \begin{array}{ll}
                m_+, & x\geq 0;\\
              -  m_- , & {\rm otherwise}.
                        \end{array}
               \right.
\label{m}
\eea
Here $f_\pm$ and $m_\pm$ are positive constants.

The case with $f_\pm=0$ and $m_+=m_-$ was considered in \cite{JR} which first showed  the existence of topological excitations or solitons in one-dimensional systems. Extension to the case with $f_\pm=0$ and $m_+\neq m_-$ is presented in \cite{shen}.

Following \cite{shen},  we assume the wavefunction in the region $x\geq 0$ to have the form 
\bea
\psi(x)=\left(
          \begin{array}{l}
             \psi_1^+\\
             \psi_2^+
             \end{array}
             \right)\,e^{-\lam_+ x},  ~~~\lam_+>0.
\eea
For $\psi(x)$ to satisfy Eq.(\ref{dirac1b}), $\lam_+$ is solved from the related secular equation.  The solution  for $\lambda_+>0$ is
\be
\lam_+=\sqrt{f_+^2 + m_+^2 -E^2}.
\ee
The two components of $\psi(x)$ are related by
\be
\psi_1^+=i\frac{\lam_+ + f_+}{E-m_+}\psi_2^+.
\ee

Similarly, for $x<0$ we take
\bea
\psi(x)=\left(
          \begin{array}{l}
             \psi_1^-\\
             \psi_2^-
             \end{array}
             \right)\,e^{\lam_- x},  ~~~\lam_->0,
\eea
with
\be
\lam_-=\sqrt{f_-^2 + m_-^2 -E^2}.
\ee
Now the two components of $\psi(x)$ are related by
\be
\psi_1^-=-i\frac{\lam_- + f_-}{E+m_-}\psi_2^-.
\ee

The wavefunction must be continuous at $x=0$. This requires $\psi^+_{1,2}=\psi^-_{1,2}$, which leads to
\be
\frac{\lam_+ + f_+}{E-m_+}= - \frac{\lam_- + f_-}{E+m_-}.
\label{match}
\ee
There exists a solution  to Eq.(\ref{match}) if 
\be
E=0, ~~ f_+=f_-=0,~~ m_+, m_-: {\rm arbitrary},
\ee
or
\be
E=0, ~~ f_+=f_-\equiv f ,~~ m_+=m_-\equiv m.
\ee

The former case has been discussed in \cite{shen}.  For the  latter case, the normalized wavefunction is
\be
\psi(x)=m\sqrt{\frac{\lam}{[(\lam+f)^2+m^2]}}\,
\left(
      \begin{array}{c}
      \frac{\lam+ f}{m}\\
      i
      \end{array}
\right)\,e^{-\lam|x|}.
\label{case2}
\ee

We note that in the limit $f\to 0$, the solution (\ref{case2}) reduces to the Jackiw-Rebbi solution given in \cite{JR}.

Also, the situation where $f\to -f$ , or $m\to -m$ can be discussed similarly.

\section{Another approach to zero energy states}

In the previous section, zero energy states were considered for the Dirac oscillator system in which $f(x), m(x)$ and $V(x)$ are all proportional to a function $W(x)$ as in Case I, or for a system whose $f(x)$ and $m(x)$ have similar functional form as in Case II. Now we look for zero energy states  of Eq.(\ref{dirac1b}) where $f(x)$ and $m(x)$ are different functions.

In terms of the component spinors Eq.(\ref{dirac1b}) reads
\be
(p_x-if)\psi_1=(E+m)\psi_2,~~~~(p_x+if)\psi_2=(E-m)\psi_1.
\ee
It is seen that the above equations are coupled and for the sake of definiteness we choose to eliminate the component $\psi_2$. The equation for $\psi_1$ reads
\be\label{1}
-\psi_1^{''}+(f^2-f^\prime)\psi_1+\f{m^\prime\psi_1^\prime+m^\prime f\psi_1}{(E+m)}=(E^2-m^2)\psi_1.
\ee
Now writing the component spinor as
\be
\psi_1=e^{\alpha(x)}\phi_1, ~~~~\alpha(x)=\f{1}{2}log(E+m),
\ee
Eq.(\ref{1}) may be written as
\be
-\f{d^2\phi_1}{dx^2}+V_1(x,E)\phi_1=E^2\phi_1,
\ee
where
\be
V_1(x,E)=f^2-f^\prime+\f{3{m^\prime}^2-2(E+m)m''}{4(E+m)^2}+\f{m^\prime}{(E+m)}f+m^2.
\ee

Let us now search for $E=0$ solutions. For $E=0$ we obtain
\be
V_1(x,0)=(f+g)^2-(f+g)^\prime+m^2,~~~~g=\f{m^\prime}{2m}.
\ee
As an example, let us take $f+g=W=\lambda \tanh x +\nu$, then $m^2=2 W'=2\lambda\, {\rm sech}^2 x$.  So we have
\begin{eqnarray}
V_1(x,0) &=&(\lam^2 + \nu^2)-\lambda(\lambda-1)\,{\rm sech}^2 x + 2\lambda\nu \tanh x,\nonumber\\
&&~~E=0.
\end{eqnarray}
Now comparing $V_1$ with the potential \cite{khare}
\bea
  &&V(x)=-A(A+1)  {\rm sech}^2 x + 2 B \tanh x, \nonumber\\
  &&~~~A, B>0, ~~B< A^2, ~~~ \nonumber\\
  && ~~E_n = -(A-n)^2 - \frac{B^2}{(A-n)^2},
  \label{khare}
  \eea 
we find
 \bea
&& A=\lam -1, ~~ B=\lam\nu, ~~\n\\
&& \lam^2 + \nu^2 =   (\lam -1-n)^2 + \frac{\lam^2\nu^2}{(\lam-1-n)^2}. \label{lam}
\eea

 Rearranging (\ref{lam}), we get
 \[
  \lam^2   -(\lam -1-n)^2 =   \left[\lam^2   -(\lam -1-n)^2 \right]\, \frac{\nu^2}{(\lam-1-n)^2} .
  \]
 So 
 \[
 \nu=\pm (\lam-1-n).
  \]
  For $B>0$, we should take $+$-sign: $ \nu= \lam-1-n >0$, i.e., $\lam>n+1$. For $A=\lam-1>0$, $\lam>1$.

The constraint $B<A^2$ leads to
\be
\lam\nu < (\lam-1)^2
\ee

or
\be
 (n-1)\lam>-1.
\ee

So
\[
n\geq 1, ~~\lam \geq n+1 > 2, ~~\nu=\lam-1-n\geq 0.
\]
For these parameters, we can have consistent solutions.

\section{Conclusion} 

In this paper we have obtained exact solutions of the generalized DO with a position dependent mass and with/without an electric field when the oscillator potential, the mass function, and the electric field are related.  In particular, we have obtained both zero as well as non-zero energy solutions. It has also been shown that for the bound states to exist the magnitude of the electric field has to be less than a critical value. Furthermore, zero energy solutions have been obtained even when the inter-relation between the oscillator and the mass function is relaxed. We believe the models considered here can be further explored with a view to obtain exact solutions with other types of mass function, e.g., periodic ones as well as studying other aspects like scattering in Dirac materials.

\acknowledgments

This work is supported in part by the Ministry of Science and Technology (MoST)
of the Republic of China under Grant MOST 106-2112-M-032-007 and MOST 107-2112-M-032-002.
PR wishes to thank MoST (R.O.C.)  and
 Tamkang University for supporting a visit during which part of the work was done.



\begin{thebibliography}{99}

\bibitem{moh} M. Moshinsky and A. Szczepaniak, J. Phys. {\bf A22} (1989) L817.

\bibitem{do1} J. Ben\'itez, R. P. Mart\'inez y Romero, H. N. N\'u\~nez-Y\'epez and A. L. Salas-Brito, Phys. Rev. Lett. {\bf 64} (1990) 1643.

\bibitem{do2} F. Domínguez-Adame and M. A. Gonz\'alez, EPL {\bf 13} (1990) 193.

\bibitem{do3} C. L. Ho and P. Roy, Ann. Phys. {\bf 312} (2004) 161. 

\bibitem{do4} C. Quesne and V. M. Tkachuk, J. Phys. {\bf A38} (2005) 1747.

\bibitem{do5} A. Bermudez, M. A. Martin-Delgado, and A. Luis, Phys. Rev. {\bf A77} (2008) 063815.

\bibitem{do6} F. M. Andrade, E. O. Silva, M. M. Ferreira Jr and E. C. Rodrigues, Phys. Lett. {\bf B731}(2014) 327.

\bibitem{jc1} P. Rozmej and R. Arvieu, J. Phys. {\bf A32} (1999) 5367.

\bibitem{jc2} A. Bermudez, M. A. Martin-Delgado and E. Solano, Phys. Rev. {\bf A76} (2007) 041801(R).

\bibitem{jc3} L. Lamata, J. Le\'on, T. Schaetz and E. Solano, Phys. Rev. Lett.
{\bf 98} (2007) 253005.

\bibitem{jc4} D. Dutta, O. Panella and P. Roy, Ann. Phys. {\bf331} (2013) 120.

\bibitem{expt} J. A. Franco-Villafa\~ne, E. Sadurni, S. Barkhofen, U. Kuhl, F. Mortessagne and T. H. Seligman, Phys. Rev. Lett. {\bf 111} (2013) 170405.

\bibitem{longhi} S. Longhi, Optics Lettrs. {\bf 35} (2010) 1302.

\bibitem{kats} K. J. A. Reijnders, D. S. Minenkov, M. I. Katsnelson, S. Yu. Dobrokhotov, 
Ann. Phys. {\bf 397} (2018) 65.

\bibitem{tkachuk} H. P. Laba and V. M. Tkachuk, Eur. Phys. J. Plus {\bf 133} (2018) 279.

\bibitem{khare} F. Cooper, A. Khare and U. Sukhatme, {\it Supersymmetry in Quantum Mechanics}, World Scientific , Singapore (2002). 

\bibitem{z1} 
J.H. Bardarson, M. Titov  and P.W. Brouwer,  Phys. Rev. Lett., {\bf 102} (2009) 226803.
\bibitem{z2} 
R.R. Hartmann, N.J. Robinson and M.E. Portnoi,  Phys. Rev. B {\bf 81} (2010) 245431.
\bibitem{z3} 
C.A. Downing, D.A. Stone and M.E. Portnoi, Phys. Rev. B { \bf 84} (2011) 155437.
\bibitem{z4} 
R.R. Hartmann, I.A. Shelykh and M.E. Portnoi, Phys. Rev. B {\bf 84} (2011) 035437.
\bibitem{z5} 
D.A. Stone, C.A. Downing and M.E. Portnoi, Phys. Rev. B {\bf 86} (2012) 075464.
\bibitem{z6} 
I.F. Herbut , Phys. Rev. Lett. {\bf 99} (2007) 206404.
\bibitem{z7} 
M. Katsnelson and M.H. Prokhorova , Phys. Rev. B {\bf  77} (2008)  205424.
\bibitem{z8} 
P. Potasz,  A.D. G\"ucl\"u  and P. Hawrylak, Phys. Rev. B {\bf 81} (2010) 033433.
\bibitem{z9} 
L. Brey and H.A. Fertig, Phys. Rev. B  {\bf 73} (2006) 235411.
\bibitem{z10}
C.-L. Ho and P. Roy, EPL {\bf 108}  (2014) 20004.
\bibitem{z11}
C.-L. Ho and P. Roy, EPL  {\bf 112} (2015) 47004.


\bibitem{JR} R. Jackiw and C. Rebbi, Phys. Rev. {\bf D 13} (1976) 3398.

\bibitem{shen} S.-Q. Shen, {\it Topological Insulators}, 2nd ed., Springer Nature, Singapore (2017).

\end{thebibliography}
\end{document}